\documentclass{epl}

\title{A lattice model of hydrophobic interactions}
\author{Daniel W. Hone\inst{1,2} \and P.A. Pincus\inst{1,3,4}}

\institute{                    
  \inst{1} Physics Department, UCSB, Santa Barbara, CA 93106\\
  \inst{2} Kavli Institute for Theoretical Physics, UCSB, Santa Barbara, CA 93106\\
  \inst{3} Materials Department and Biomolecular Science and Engineering, 
UCSB, Santa Barbara, CA 93106\\
  \inst{4} Physics Department, KAIST, Daejon, South Korea
}

\begin{document}

\maketitle

\begin{abstract}
Hydrogen bonding is modeled in terms of virtual exchange of protons between water molecules.  A simple lattice model is analyzed, using ideas and techniques from the theory of correlated electrons in metals. Reasonable parameters reproduce observed magnitudes and temperature dependence of the hydrophobic interaction between substitutional impurities and water within this lattice.
\end{abstract}

\section{Introduction}
Hydrogen bonding \cite{chandler,dillrev,dillbook}    is responsible for the cohesion of water and is intimately involved in solvation.  Protein secondary structures such as $\alpha$ helices and $\beta$ sheets are engendered by intramolecular H bonds that are stabilized in competition with water bonding.  Such systems suggest that hydrogen bond correlations are relevant for their understanding.  The purpose of this communication is to propose a simple toy lattice model \cite{widom} that allows consideration of proton correlations.  We are motivated by the Hubbard model \cite{rasetti}  for correlated electron systems, which has illuminated the behavior of  transition metals, high temperature superconductors, heavy Fermion conductors, etc.  Here we introduce for simplicity the most primitive version of our H-bond model, with molecular sites restricted to a simple cubic lattice.  The phenomena of concern here involve short range effects, where the long range disorder of liquids is not paramount.

The molecular structure of the water molecule is closely related to the unusually large electronic polarizability of the O$^{--}$ ion. Pure coulomb considerations with a rigid oxygen atom would place the protons in a water molecule directly opposite one another, whereas the low energy configuration, due to the oxygen polarizability, locates them at a relative angle of $104$ degrees.  We simulate this feature in a model which is simpler geometrically, with possible proton locations only at the six cartesian positions $\pm x$, $\pm y$, $\pm z$ of a simple cubic lattice.  These positions, which we call basins, are located in the vicinity of the outer edge of the oxygen ion electronic cloud.  The basins are meant to be local minima of the potential energy for the protons and would be determined by a balance between attraction to the nearest O$^{--}$  and the crystalline electric fields associated with the nearby oxygens.  We take the binding energy of a single proton in a basin associated with one of the O$^{--}$ ion as the zero of potential energy. We treat the occupied basins as quantum states and the protons as spinless fermions, so that any basin can be occupied by at most one proton. The potential energy cost for placing the second proton on an oxygen site already containing one proton will be $V$ if the second site is  perpendicular to the first (near neighbor), whereas the cost for putting the second opposite the first (next near neighbor) will be $U > V$.  We consider a simple cubic lattice of such oxygen sites.  Consider two neighboring water molecules (two protons on each) on that lattice in the x-direction (see Fig. 1). 
\begin{figure}[h]
\begin{center}
\includegraphics[scale=0.6]{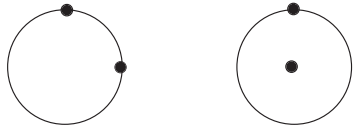}
\caption{ Neighboring $H_2O$  molecules.  The occupied proton basins are denoted by the small dark filled circles; the circle at the center on the right indicates the basin above the plane of the figure.}
\label{fig1}
\end{center}
\end{figure}
 If there is a proton on the left molecule in the $+x$ site, and the $-x$ site on the right hand molecule is empty, then the proton can tunnel from the left to the right molecule, leaving a hydroxyl ion OH$^-$ on the left and creating a hydronium ion H$_3$O$^+$ on the right.  The minimum energy cost for doing this is $V$, the value appropriate to the configuration illustrated in Fig. 1. We denote the tunneling matrix element by $t$.  The notation has intentionally been chosen to draw attention to the close analogy of this with the well studied Hubbard model of correlated electron systems. The pH of water at room temperature suggests, by the law of mass action, a value of $V \approx 32k_BT$, or about 0.8 eV.  The virtual hop and return of a proton between neighboring molecules forms the hydrogen bond within the model, with a strength given in lowest order perturbation theory by $-t^2/V$, suggesting $t \approx (5-10)k_BT$ to yield the known strength of that bond at room temperature.  The value of $U$ will be greater than, but of the same order as $V$.  Transitions, real or virtual, involving energetic cost $U$ will be thermodynamically suppressed relative to those involving a cost $V$.  Because it greatly simplifies the algebra without significant alteration to the physical conclusions, we will henceforth take $U$ as effectively infinite, so that all corresponding configurations can be neglected.  We note that we are also neglecting coulomb energies associated with ionic configurations.  This may be of some consequence in the detailed physical behavior, but the number of real ions at temperatures of interest is extremely small.

This picture also suggests a source of hydrophobic interactions \cite{chandler,dillrev,dillbook,widom,tanford,pratt,zichi,naim}.  Consider replacing two of the water molecules by others without mobile protons, such as organic molecules.  If these impurities are well separated, they each eliminate six hydrogen bonds, whereas if the impurities are near neighbors on the lattice, they eliminate only 11 such bonds.  Energetics promotes clustering of the impurities, equivalent to a hydrophobic interaction.  Of course, there are entropic effects, as well.  Below we calculate the free energy, to second order in the hopping $t$, in order to estimate the size of this hydrophobic effect. The lattice model is, of course, most explicitly relevant to systems like clathrates or ice.  The neglect of global reorganization of water in liquids overestimates the hydrophobic interaction there.

We write the Hamiltonian $H$ (including the chemical energy term $-\mu N$) as the sum of a zeroth order term $H_0$, with no hopping ($t=0$), plus the hopping perturbation $H_t$.   We will calculate the free energy from the log of the grand partition function,
\begin{equation}
Z = {\rm Tr} e^{- H} ,
\label{partition}
\end{equation}
with a corresponding definition for the unperturbed grand partition function $Z_0$.  We have taken units of energy as $k_BT$ here for simplicity of notation. The free Hamiltonian can be written in terms of the parameters defined above as
\begin{equation}
H_0 = \frac{V}{2} \sum_{i,\alpha\neq\gamma} (n_{i\alpha} + n_{i\bar\alpha})(n_{i\gamma} + n_{i\bar\gamma}) + U\sum_{i,\alpha} n_{i\alpha}n_{i\bar\alpha} - \mu\sum_{i,\alpha} (n_{i\alpha} + n_{i\bar\alpha} )
\label{h0}
\end{equation}
Here we have labeled the proton sites by 2 indices.  The roman index $i$ indicates the lattice site of the attached oxygen, and the greek index $\alpha$ labels the location on the oxygen: $\alpha = x,y,z$ and $\bar\alpha = -\alpha$. The  proton number operator on the site ${i,\alpha}$ is denoted by $n_{i,\alpha} = c^{\dagger}_{i,\alpha}c_{i,\alpha}$ .  The hopping Hamiltonian is
\begin{equation}
H_t = t \sum_{i,\alpha} \left[ c^{\dagger}_{i+\alpha,\bar\alpha} c_{i,\alpha} + {\rm H.c.} \right].
\label{ht}
\end{equation}
A standard cumulant expansion gives as the second order correction to the free energy,
\begin{equation}
\ln \frac{Z}{Z_0} \approx \int_0^1 d\lambda \int_0^{\lambda'}\langle H_t(\lambda') H_t(\lambda') \rangle_0,
\label{pertf}
\end{equation}
where we have used the standard notation for operators in the interaction picture,
\begin{equation}
H_t(\lambda) = e^{\lambda H_0} H_t e^{-\lambda H_0},
\end{equation}
and the angle brackets with subscript zero imply a thermal average determined by the unperturbed Hamiltonian,
\begin{equation}
\langle A \rangle_0 \equiv  (Z_0)^{-1} {\rm Tr} A e^{-H_0}
\label{trace0}
\end{equation}

\section{Non-interacting free energy}

Since $H_0$ is the sum over independent oxygen site Hamiltonians, the corresponding partition function is the product of $N$ identical site terms for a lattice of $N$ sites: $Z_0 = z_0^N$.  Moreover, we note that each of the terms in $H_0$ depends only on $N_{\alpha} = n_{\alpha} + n_{\bar\alpha}$, which simplifies the enumeration of terms.  In principle, we should sum over proton number occupation on a site ranging from 0 to 6, but the only terms which will play any important role are the obvious physical ones, namely those with numbers 1, 2, and 3 (hydroxyl ion, water, and hydronium, respectively).  Then we find
\begin{equation}
z_0 \approx 6e^{\mu} + 12e^{2\mu - V}  + 8e^{3\mu -3V}
\label{z0}
\end{equation}
As always, the chemical potential $\mu$ is determined by specifying the average number of protons, which here is 2 per site: 
\begin{equation}
2 = \frac{\partial}{\partial\mu} \ln z_0,
\end{equation}
which gives
\begin{equation}
e^{\mu_0} \equiv \zeta_0 \approx \frac{\sqrt 3}{2}e^{3V/2},
\label{fugacity0}
\end{equation}
and the free energy
\begin{equation}
F_0/N = \mu_0 - T\ln z_0.
\label{freeenergy0}
\end{equation}

\section{Interaction energy; high temperature}

We must next calculate the second order perturbation correction to the free energy, given in Eq. (\ref{pertf}).  Again there are many equivalent terms. Each nonvanishing term in the trace (see Eq.(\ref{trace0})) involves the hop of a proton to a neighboring oxygen site and its subsequent return to the initial site.  Each of the cubic directions is equivalent, so we need to consider only a single pair of oxygens, with only a single pair of sites between which the proton can hop.  Every term in the trace depends on $\lambda$ and $\lambda'$ in the form $\exp[(\lambda'-\lambda)W] $, which gives for the integrals over those variables in (\ref{pertf}),
\begin{equation}
\int_0^1 d\lambda \int_0^{\lambda} d\lambda' e^{(\lambda'-\lambda)W} = \frac{e^{-W}-1}{W^2} + \frac{1}{W}.
\label{lambdaint}
\end{equation}
The energies $W$ are positive or negative multiples of $V$, which is large compared to unity in magnitude at any temperature of interest to us.  If $W>0$, then the result is approximately $1/W - 1/W^2$.  If $W<0$, then it is approximately $e^{-W}/W^2$. If we neglect all terms exponentially small relative to the leading one,  then we find the simple result,
\begin{equation}
\ln \frac{Z}{Z_0} \approx   \frac{6Nt^2}{z_0^2}{\rm Tr}e^{-(H_0^i+H_0^{i+x})}\int_0^1 d \lambda \int_0^{\lambda} d \lambda' \langle H_t^i(\lambda)H_t^{i+x}(\lambda')\rangle_0 \approx \frac{2Nt^2}{3}\left(\frac{1}{V} - \frac{1}{V^2} \right).
\label{leadingterm}
\end{equation}
This represents the high temperature limit, in the sense that we require $t^2/V\ll 1$ (though we still assume $V\gg 1$), restricting the range of validity for physically realistic parameters to twice room temperature or more.  We can include the leading corrections which depend on the chemical potential, which will be necessary to determine the free energy.  The result for the partition function with all terms involving site proton occupancies of 1, 2, or 3 only, is:
\begin{equation}
\ln \frac{Z}{Z_0} \approx 
N \frac{12 t^2 \left \{ 2\left(\frac{1}{ V}-\frac{1}{ V^2}\right) +e^{-(\mu -V)} \right\} }  {\left[ 6+3e^{-(\mu-V)} + 4e^{(\mu-2V)}\right]^2 } .
\label{relativez}
\end{equation}
With the fugacity given approximately by (\ref{fugacity0}) we see that the limit (\ref{leadingterm}) is given by the first factors in numerator and denominator here, with corrections of order $e^{-V/2}$ multiplied by 1, $V$, or $V^2$. At constant volume the (Helmholtz) free energy is given by $F = \bar N\mu  -T\ln Z$,
with the chemical potential $\mu$ determined by
\begin{equation}
\bar N = 2N = \frac{\partial}{\partial\mu} \ln Z.
\label{mu}
\end{equation}
 In the correction terms of order $t^2$ we can take $\mu\approx\mu_0$, as given by (\ref{fugacity0}).  Then we find for the fugacity $\zeta = e^{\mu}$ the approximate result
\begin{equation}
\zeta^2 \approx \frac{3+2t^2}{4}e^{3 V} .
\label{fugacity}
\end{equation}

We can immediately find the free energy (we re-insert all factors of the temperature $T$ now
for clarity) as
\begin{eqnarray}
F &=& 2N\mu - T\ln Z    \nonumber\\
&\approx&  N\left[ V  - \frac{2t^2}{3}(1/V - T/V^2)- 
\frac{2T}{\sqrt 3}e^{-\beta V/2}-T\ln(12) \right]   \qquad ( \beta t\ll 1; \beta V\gg 1).
\end{eqnarray}
The first two terms are the average energy up to second order corrections in the hopping, and the last term is the familiar form of the entropy in the high temperature limit for the set of 12 states on each site where the two protons are near neighbors.  Of course, we can trust this expansion only down to temperatures of order $t^2/V$, well above the temperatures of physical interest for water.   

But in this high temperature regime we can readily estimate the hydrophobic interaction as described above, by looking at the correction to the free energy due to the hopping term $t$ in the Hamiltonian, which is what is eliminated locally when a water molecule is replaced by an organic or other molecule without mobile hydrogen atoms.  We have
\begin{equation}
\delta F = F-F_0 = \bar N (\mu - \mu_0) - T \ln(Z/Z_0) = NT\ln(\zeta/\zeta_0)^2 - T \ln(Z/Z_0)
\label{deltafdef}
\end{equation}
We have expressions for each of these terms, using (\ref{fugacity0}), (\ref{fugacity}), and (\ref{relativez}).  If we keep only the leading exponential order of temperature dependence, we find
\begin{equation}
\delta F \approx -\frac{2Nt^2}{3T}\left[ 1-\frac{T}{V}+\frac{T^2}{V^2} - \frac{\sqrt 3}{4} e^{-\beta V/2} \right]
\label{deltaf}
\end{equation}
The hydrophobic interaction, then, is the difference in free energy between two impurities well separated and the same two as near neighbors, with one fewer virtual hopping opportunity excluded.  We can read off this attractive interaction directly from (\ref{deltaf}):
\begin{equation}
V_{hydro} =  -\frac{2t^2}{3T}\left[ 1-\frac{T}{V}+\frac{T^2}{V^2} - \frac{\sqrt 3}{4} e^{-\beta V/2} \right]
\end{equation}

\section{Low temperature limit}

We can also estimate the ground and low lying excited state energies to comparable accuracy to examine the low temperature limit and thereby to obtain an idea of the full temperature dependence of the interaction free energy which, as described above, suggests an origin of the hydrophobic interaction.  

In the absence of hopping, $t = 0$, it is clear that the ground state of the model consists of two protons in nearest neighbor sites on each oxygen.  There is complete degeneracy with respect to the location of the independent pairs on each site.  To second order in the hopping the most favorable situation corresponds to each hydrogen being able to tunnel virtually to a neighboring oxygen and then back, leaving in the virtual intermediate state a hydroxyl ion and  neighboring hydronium ion, with the minimum possible extra intermediate state energy, namely $V$.  In fact, we can construct such an arrangement (see Fig. 2).  
\begin{figure}[h]
\begin{center}
\includegraphics[scale=0.4]{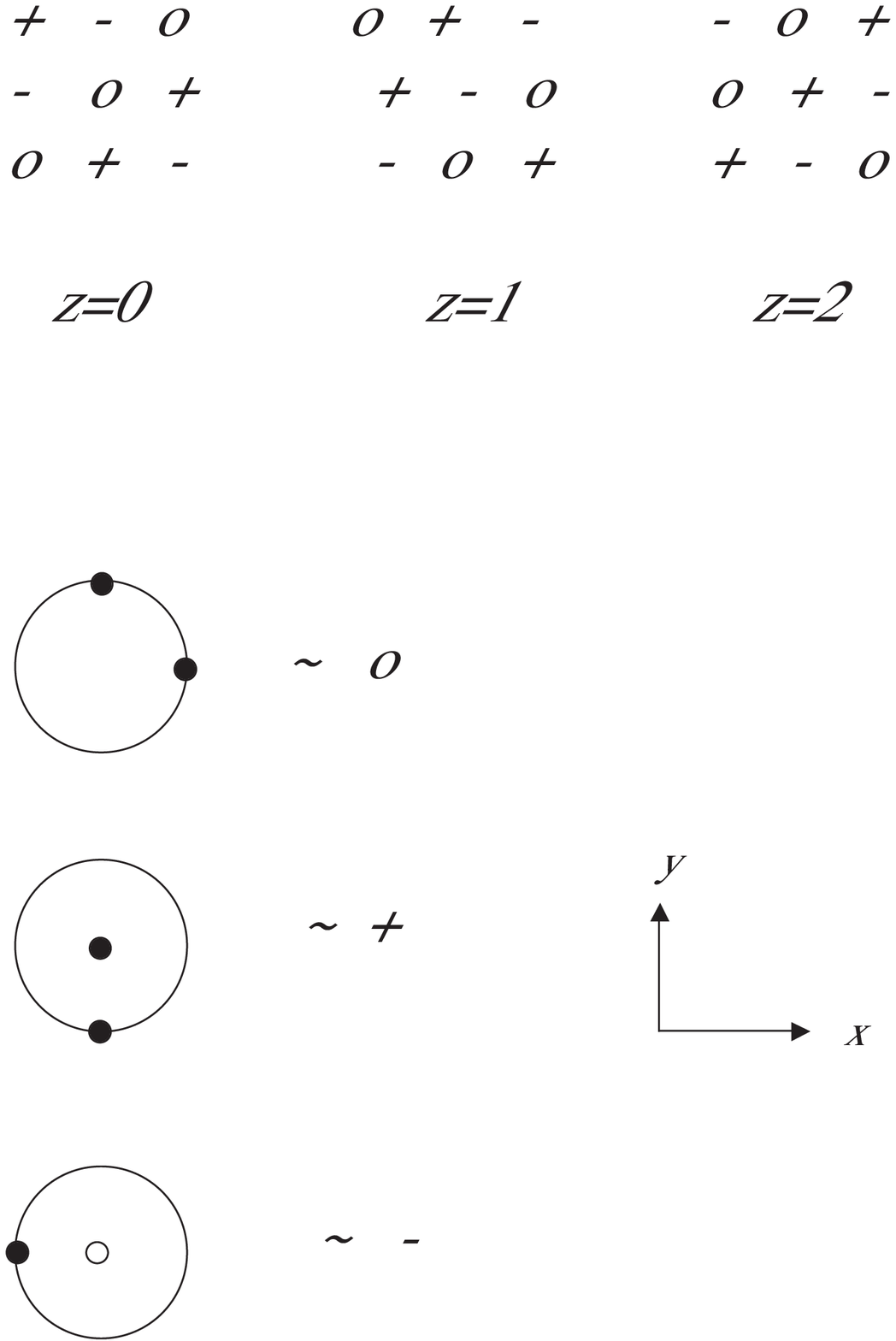}
\caption{ Ground state configuration in the $z=0$ plane.  Open small circle lies below the plane of the figure.  Successive planes above this one can be generated by shifting each row of the plane below down one row.}
\label{default}
\end{center}
\end{figure}
We take as a basis set for the doubly occupied sites the three near neighbor pairs $xy$, $\bar yz$, and $\bar x\bar z$, which we denote 0, +, and - for notational convenience.  At the origin we place a 0 molecule, at the position (1,0,0) a + molecule, and at (0,1,0) a - molecule.   This configuration is repeated, with the basis vectors 110, $\bar 1$20, 0$\bar 1$1, giving three successive layers along the z-axis as shown in Fig. 2.  Futher layers simply repeat this pattern.  It is easy to check that this allows for every proton to participate in a tunneling process in which the intermediate hydronium has a proton in each of the three coordinate locations, so the intermediate energy is indeed $V$ in every case.   There is clearly some degeneracy, in addition to the obvious states arising from translational and rotational  symmetry.  That can be removed in higher order, but to second order in $t$ this is the ground state energy of interaction, $-2Nt^2/V$.  

We can further estimate the low temperature thermodynamics by examining the low lying excited states. There are the above-mentioned states which differ from the ground state only at higher order in the hopping $t$.  But the number of these is not extensive (of order the number of sites $N$).  There are, however, an extensive number of states in which a single site is occupied by a pair of near neighbor protons in such a way that the intermediate energy for hopping of one of them is $U$ (which we have been taking infinite, or at least large enough to ignore) instead of $V$, and the interaction energy associated with the hopping of that proton is now $-t^2/U$ (zero, to be consistent with our earlier approximation for $U$) instead of $-t^2/V$.  There are $2N$ of these lowest excited states, obtained by taking a proton located in the direction $w$ on an oxygen site in the ground state and moving it to the location $-w$, with $w=x,y,$ or $z$.  Thus there is an energy gap to the lowest (thermodynamically relevant) excited states.  Since the low lying excited states are non-ionic (always with two protons on each site), we can calculate the free energy at low temperatures within the canonical ensemble:
\begin{equation}
F = -T\ln Z \approx N\left[ V - \frac{2t^2}{V} - T\ln\left( 1+6e^{-\beta t^2 /V} \right)\right] \qquad 
(\beta t \gg 1),
\end{equation}
which decreases only exponentially slowly at low temperatures.  Again, the hydrophobic repulsion corresponds to the blocking of one fewer virtual hopping sites when the impurities are near neighbors, so that interaction in this low temperature limit is
\begin{equation}
V_{hydro} \approx -2t^2/V -T\ln\left( 1+6e^{-\beta t^2 /V} \right)
\end{equation} 

We plot the behavior of the hydrophobic interaction as a function of temperature in Fig. 3, using the limiting low and high temperature expressions above.  
\begin{figure}[htbp]
\onefigure[scale=0.6]{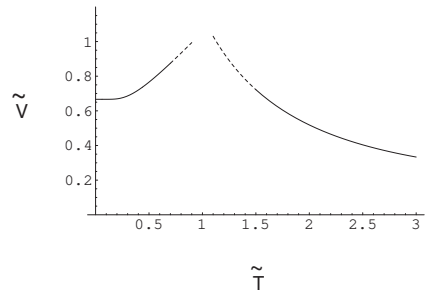}
\caption{The dimensionless hydrophobic attraction $-\tilde V=V_{hydro}(V/t^2)$  as a function of dimensionless temperature $\tilde T\equiv TV/t^2$ , with $(V/t)^2=6$  (see Eq. (\ref{vtilde})).}
\label{fig3}
\end{figure}
It is convenient to scale both the interaction and temperature by the characteristic energy of the model $t^2/V$: $\tilde V_{hydro}\equiv V_{hydro}V/t^2$, and $\tilde T\equiv TV/t^2$.  Then
\begin{eqnarray}
\tilde V = \cases{2+\tilde T\ln\left(1+6e^{-1/\tilde T}\right) & $(\tilde T\ll 1)$;\cr (2/3\tilde T)(V/t)^2 & $(\tilde T\gg 1)$. \cr}
\label{vtilde}
\end{eqnarray}
The result contains one factor of the parameter $(V/t)^2$, which is of order 3 to 10.  We have taken it as 6 for purposes of the plot, but the detailed result does depend on the specific value chosen.  The results for the model do require an interpolation between the limiting curves, but we can conclude  the hydrophobic interaction grows at low temperature to some ultimate maximum somewhat above room temperature, and then decreases. This suggests an increase in the hydrophobic interaction in the range of a few percent up to ten percent or so over a temperature change of, say 50 kelvin at temperatures where water is liquid, with values several times $k_BT$, which is the correct sign and magnitude for what is observed.

In summary, we have introduced a primitive Hubbard-like model to describe
hydrogen bonding networks, that has the ability to rationalize the hydrophobic
interaction in terms of disruption of this network.  Natural extensions include
hydration of ionic impurities, hydronium mobility, and competitive hydrogen
bonding situations that occur in solubilization of polypeptides and
polyethylene glycol.

\acknowledgments
We would like to acknowledge very helpful discussions with D. Scalapino, N. Gov,
K. Dill, Y. S. Jho, and J. Eckert.  Supported in part by NSF-DMR 0503347 and the US-Israel
Binational Science Foundation.  PP would like to thank the
Aspen Center for Physics for their hospitality.

\end{document}